\documentclass[12pt]{article}
\usepackage{setspace}
\usepackage{amssymb, amsfonts, amsmath}
\usepackage{graphicx}
\usepackage{cite}
\usepackage[margin=1in]{geometry}
\usepackage{authblk}

\begin{document}

\title{Freezing on a Sphere}

\author{Rodrigo E. Guerra*}
\author{Colm P. Kelleher*}
\author{Andrew D. Hollingsworth}
\author{Paul M. Chaikin}
\affil{Center for Soft Matter Research\\
New York University, New York, NY 10003}

\date{\today}

\maketitle
\doublespacing

The best understood crystal ordering transition is that of two dimensional freezing, which proceeds by the rapid eradication of lattice defects as the temperature is lowered below a critical threshold~\cite{StrandburgRMP1988, HalperinNelsonPRL1979, KosterlitzThoulessJPhys1973, GasserChemPhysChem2010, YoungPRB1979, NelsonDefectsBook}.
But crystals that assemble on closed surfaces are required by topology to have a minimum number of lattice defects, called disclinations, that act as conserved topological charges --- consider the 12 pentagons on a soccer ball or the 12 pentamers in a viral capsid~\cite{CasparKlugQB1962, simmons_fundamental_1970}.
Moreover, crystals assembled on curved surfaces can spontaneously develop additional defects to alleviate the stress imposed by the curvature~\cite{IrvineNatMat2012, IrvineNature2010, turner_vortices_2010}.
How then can we have crystallization on a sphere, the simplest curved surface where it is impossible to eliminate these defects?
Here we show that freezing on a sphere proceeds by the formation of a single, encompassing ``continent", which forces defects into 12 isolated ``seas" with the same icosahedral symmetry as soccer balls and viruses.
We use this broken symmetry --- aligning the vertices of an icosahedron with the defect seas and unfolding the faces onto a plane --- to construct a new order parameter that reveals the underlying long-range orientational order of the lattice.
These results further our understanding of the thermodynamic and mechanical properties of naturally occurring structures, such as viral capsids, lipid vesicles, and bacterial s-layers~\cite{CasparKlugQB1962, sleytr_s-layers2014}, and show that the spontaneous sequestration and organization of defects can produce mechanical and dynamical inhomogeneieties in otherwise homogeneous materials.


\newpage


The remarkable predictions of the Kosterlitz-Thouless-Halperin-Nelson-Young (KTHNY) ~\cite{HalperinNelsonPRL1979, KosterlitzThoulessJPhys1973, YoungPRB1979, NelsonDefectsBook} theory of melting on a plane have been verified both by experiment and simulation in systems as diverse as free electrons on liquid He surfaces~\cite{GrimesAdamsPRL1979, StanDahmPRB1989}, magnetic flux vortices in superconductors~\cite{DodgsonMoorePRB1997, guillamon_direct_2009}, and monolayers of paramagnetic colloidal particles interacting through induced magnetic dipoles~\cite{SkjeltorpJAP1984, LinPRE2006, DillmannJPCM2012, ZahnPRL1999, ZahnPRL2000, DeutschlanderPRL2014}.
This last system is particularly well characterized --- its entire phase behavior can be reduced to a 1D phase diagram parameterized by a single, dimensionless interaction parameter, $\Gamma$, defined as the ratio of neighbor magnetic dipole and thermal energies,
\begin{equation}
\label{gammadef}
\Gamma\! =\! \frac{A\,(\pi\rho)^{3/2}}{k_B T},
\end{equation}
where $A$ quantifies the magnitude of the dipolar pair-potential, $U(r)=A\,r^{-3}$, $\rho$ is the number density of particles, $k_B$ is Boltzmann's constant, and $T$ is the absolute temperature.
Consistent with this definition, the liquid-crystalline hexatic phase that separates the isotropic fluid from the crystalline solid occupies a narrow window of $\Gamma$ values, $67 \lesssim\!\Gamma\!\lesssim 70$.

\begin{figure*}
\center
\includegraphics[width=15cm]{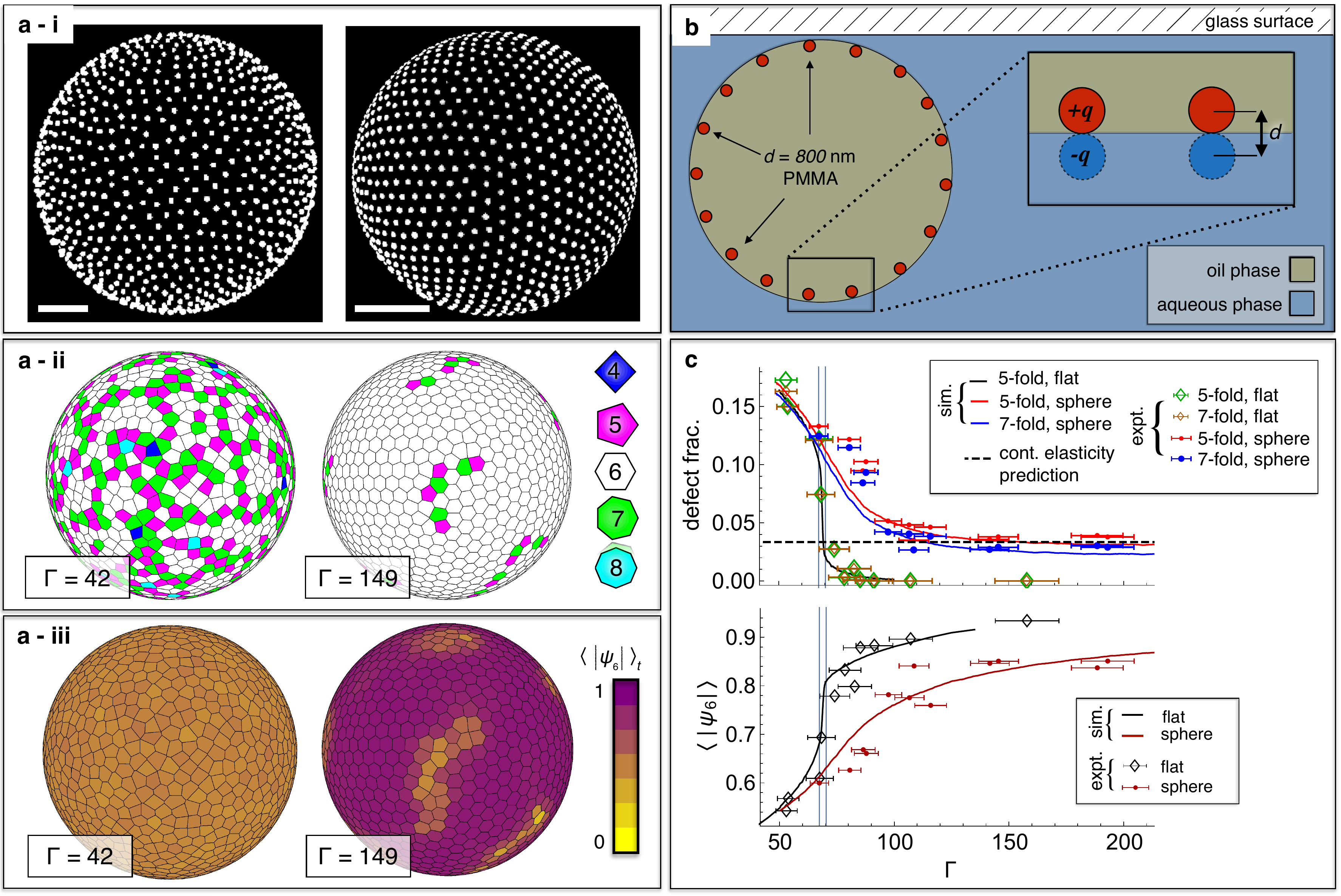}
\caption{\textbf{Defects and order in spherical assemblies} (a) i) Particles adsorbed on a sphere at low density ($\Gamma\!=\!42$, left) appear disordered.  More densely packed particles ($\Gamma\!=\!149$, right) exhibit hexagonal order and lattice lines characteristic of flat crystals.  Scale bars, 20$\,\mu$m. ii) Lattice defects uniformly cover the disordered sample, whereas the more ordered sphere is covered mostly by 6-coordinated particles. iii) Values of the time-averaged bond-orientation order parameter, $\langle|\psi_{6}(\vec{r})|\rangle_{t}$ (see text), of particles on the disordered sphere are much more uniform than those on the higher-$\Gamma$ sphere. (b) Image charges generated by the free ions in the aqueous fluid attract hydrophobic, charged particles to the oil-aqueous interface and induce dipolar repulsive forces between them. (c) Measurements of defect fractions and $\langle|\psi_{6}(\vec{r})|\rangle$ for spheres decorated with $N\!\approx\!1500$ particles and for similar particles on flat surfaces. Flat surface packings eliminate nearly all defects and attain nearly perfectly orientational order over a narrow range of $\Gamma$ values ($67 \!\lesssim\!\Gamma\!\lesssim\! 70$, shaded). For spherical packings, the improvement in local order and annihilation of defects with increasing $\Gamma$ is dramatically hindered, with a finite concentration of defects expected even in the limit $\Gamma\to\infty$ (dashed horizontal line).
\label{fig:fig1}}
\end{figure*}

Our system of charged, hydrophobic microspheres adsorbed at oil-aqueous interfaces by image charge forces is accurately described by the same interaction potential (Fig.~\ref{fig:fig1}b)~\cite{KelleherPRE2015, KelleherPRE2017, Leunissen2007, LeunissenPCCP2007}.
Yet, unlike any system controlled by magnetic fields, these particles can uniformly decorate the surfaces of spheres, or other curved surfaces, without affecting the form of their interparticle forces.
However, the topology of the sphere fundamentally changes the KTHNY picture of ordering by elimination of defects: as an excess of 12 positively charged disclinations (particles with pentagonal Voronoi cells, 5's) is required~\cite{CasparKlugQB1962, simmons_fundamental_1970}.
Furthermore, it has been shown that as the size of the system increases (at constant density) it becomes \textit{energetically} favorable to decorate these 12 disclinations with dislocations (topologically neutral, 5\,--\,7 disclination pairs) organized into linear structures called ``scars"~\cite{GarridoPRB1997, DodgsonMoorePRB1997, BowickPRL2002, BowickPRB2000, BowickPRB2006, BauschScience2003, LipowskyNature2005}.
The number of dislocations --- proportional to the length of the scars --- grows as $R/a$, where $R$ is the radius of the sphere and $a$ is the interparticle distance.
Consequently, since dislocations are mobile defects that destroy both crystallinity and rigidity~\cite{HirthAndLothe, NelsonDefectsBook}, it is natural to ask whether crystallization of particles confined to the surface of a sphere is possible at all: or if the proliferation of defects leads to liquid or glassy phases even for the strongest interactions (highest $\Gamma$).

Confocal micrographs (Fig.~\ref{fig:fig1}(a-i)) show clear differences between droplets with high and low $\Gamma$-values (for methods see Supplementary Information, SI \S1).
First, Voronoi tessellations of these surfaces show that disclinations, particles with topological charge equal to 6 minus their coordination number, densely and uniformly cover the liquid-like sample, but are much rarer in the more ordered sample and are clustered in scars (Fig.~\ref{fig:fig1}(a-ii)).
A similar pattern is shown by the 2D bond-orientational order parameter that measures the orientation and degree of hexagonal order around each particle~\cite{NelsonDefectsBook}:
\begin{equation}
\label{psi6def}
\psi_6(r_i,t)\!=\! \frac{1}{N_i}\sum^{N_i}_{j=0}e^{6\,i\,\theta_{ij}(t)},
\end{equation}
where $N_i$ is the coordination number of the $i$-th particle, and $\theta_{ij}(t)$ is the angle between the bond connecting particle $i$ to its $j$-th neighbor and an arbitrary reference axis.
Like the distribution of topological defects (Fig.~\ref{fig:fig1}(a-ii)), the time-averaged $|\psi_6(r_i,t)|$ field, $\langle|\psi_6|\rangle_{t}$, is homogeneous over the entire sphere, as shown in Fig.~\ref{fig:fig1}(a-iii).
For the denser samples, $\langle|\psi_6|\rangle_t$ is spatially heterogeneous --- the majority of particles sit in a locally hexagonal environment for the entire duration of observation, while a small number of particles, in or near the scars, have very low $\langle|\psi_6|\rangle_t$ values.

KTHNY theory postulates that the quality of this orientational order is intimately related to the distribution of topological defects~\cite{NelsonDefectsBook}.
Indeed, in flat space the number of topological defects drops precipitously over a narrow range of $\Gamma$ values centered around $\Gamma=70$, while $\langle|\psi_6|\rangle$ increases rapidly~\cite{LinPRE2006, DillmannJPCM2012}.
By contrast, while the decimation of defects on spheres with $N\!\approx\!1500$ particles is accompanied by increasing values of $\langle|\psi_6|\rangle$, there is no obvious trace of the sharp transition that occurs in the flat layers;
moreover, the number of topological defects reaches a plateau value much larger than the 12 required by topology (Fig.~\ref{fig:fig1}c).
In this $\Gamma\!\to\!\infty$ limit, the number and clustering of defects has been predicted using continuum elasticity~\cite{BowickPRB2006, BauschScience2003}.

\begin{figure*}
\center
\includegraphics[width=15cm]{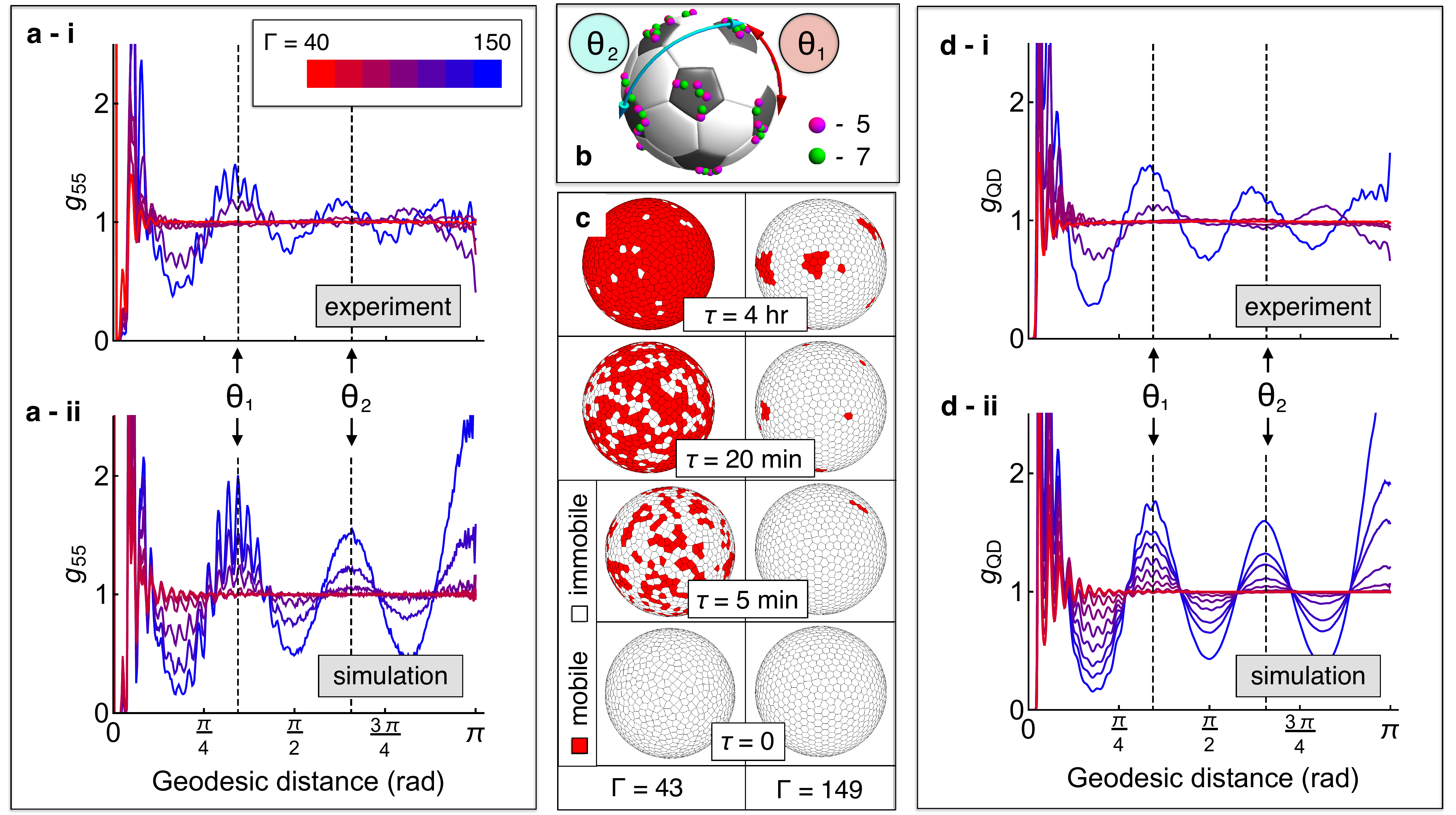}
\caption{\textbf{Emergence of icosahedrally coordinated defect ``seas''} (a) Pair correlation of 5-fold coordinated disclinations, $g_{55}$, for experimental (i) and simulated (ii) spheres decorated with $N\!\approx\!1500$ particles. Disordered spheres are uniformly randomly covered with lattice defects, leading to a flat and featureless $g_{55}$. As order builds so does a coherent defect-density modulation consistent with their segregation into icosahedrally coordinated ``seas". (b) Rendering of experimentally determined lattice defect positions over the surface of a suitably oriented soccer ball makes the icosahedral ordering apparent. (c) Distribution of ``mobile" particles (see text) over the surface of spheres with $\Gamma\!=\!42$ and $\Gamma\!=\!149$. ``Mobile" particles appear randomly over the surface of the disordered sphere and quickly cover its surface. Particles on the more ordered sphere have far lower absolute mobilities, and the ``mobile" particles that appear are segregated into isolated ``seas". (d) Pair cross-correlations of all defects and ``mobile" particles for the same spheres shown in (a) show similar ordering to $g_{55}$, demonstrating the spatial coincidence of defects and mobility. \label{fig:fig2}}
\end{figure*}

Yet, while these globally averaged quantities reflect the increasing order of particles with increasing $\Gamma$, they do not reflect the clustering of defects that is evident in the micrographs.
To better understand this organization we compute $g_{55}(r)$, the pair correlation between 5-fold defects (SI \S4), for experimental and simulated~\cite{hoomd1, hoomd2, leimkuhler_efficient_2016,PrestipinoPhysicaA1992,PrestipinoPhysicaA1993} particle configurations over a wide range of $\Gamma$.
We find that $g_{55}(r)$ is flat for $r\!\gtrsim\!a$ and $\Gamma<70$ --- consistent with a random distribution of defects.
However, for $\Gamma>70$, we find a peak in $g_{55}(r)$ at distances $r\!\gtrsim\!a$ that grows and widens with increasing $\Gamma$ --- indicating the condensation of defects into scars.
More interestingly, additional peaks in $g_{55}(r)$ appear for values of $r$ that correspond to the geodesic distances between the vertices of an icosahedron (Fig.~\ref{fig:fig2}a), making it possible to draw a soccer ball on the sphere such that most of the defects lie inside its pentagons (Fig.~\ref{fig:fig2}b).
Nevertheless, this icosahedral ordering of defects develops gradually, and it is difficult to unambiguously identify isolated defect clusters until $\Gamma \gtrsim 120$.

To investigate the effects of defect segregation on the dynamics of our system we adopt a tool used to study glasses, labeling particles that move more than a distance $\lambda^*$ over a time $\tau^*$ as ``mobile'', $Q=1$, and those that do not as ``caged", $Q=0$ (SI \S6)~\cite{parisi_short-time_1997, glotzer_time-dependent_2000}.
At large values of $\Gamma$, we observe clustering of mobile particles reminiscent of the clustering of topological defects (Fig.~\ref{fig:fig2}c).
To compare these we compute $g_{QD}(r)$, the pair correlation function between mobile particles ($Q$) and particles with any coordination number other than six ($D$), and find that it is almost identical to $g_{55}(r)$: confirming that particle mobility is strongly heterogeneous and becomes confined to the same icosahedrally coordinated ``seas'' that contain the excess lattice defects.

\begin{figure*}
\center
\includegraphics[width=15cm]{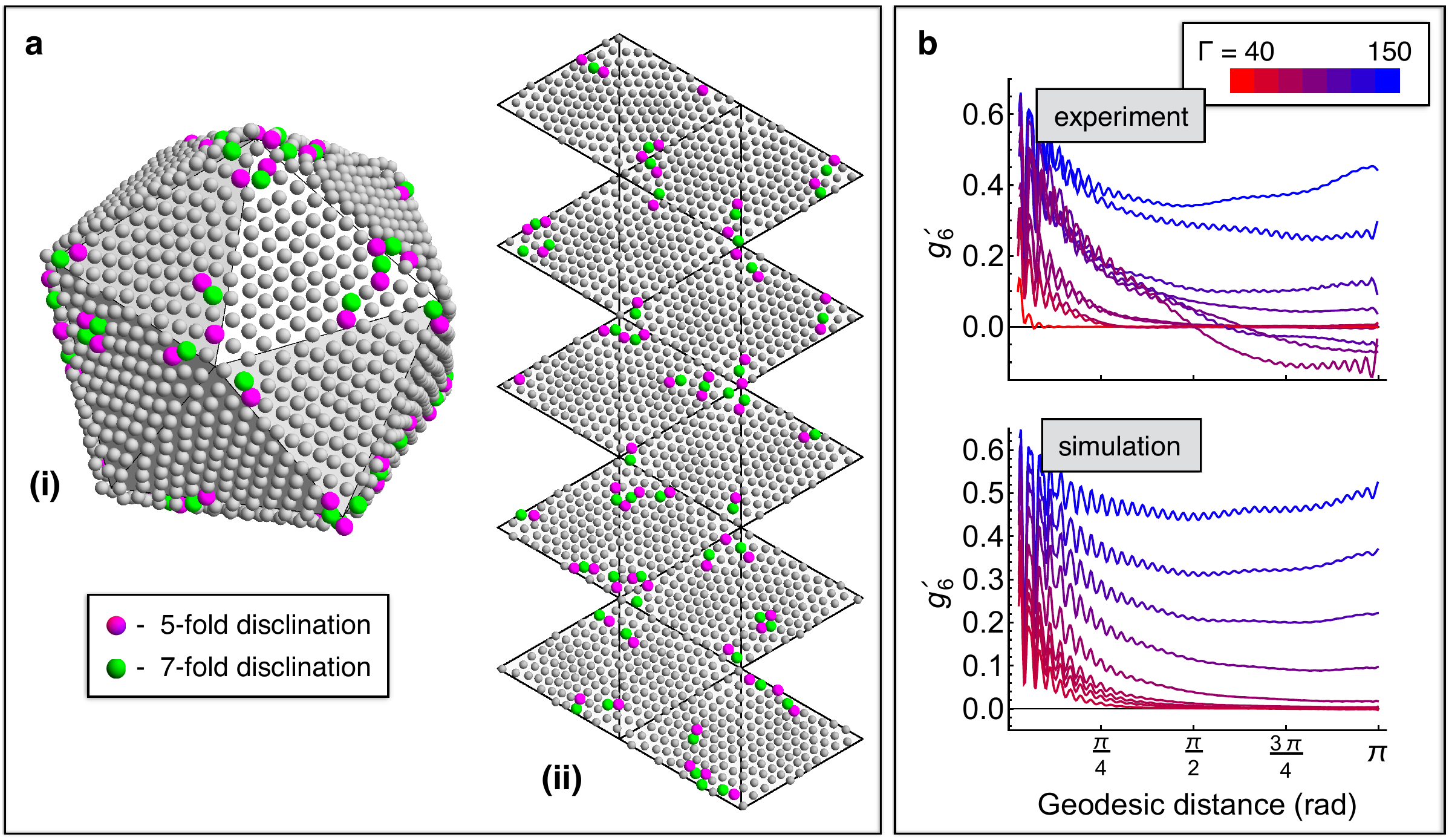}
\caption{\textbf{Icosahedral map projections and long-range order}
(a)(i) A reference icosahedron is oriented by fitting its vertices to the positions of the defect particles. (ii) Unfolding this icosahedron reveals long-range order hidden by the extra disclinations. (b) Orientational correlation functions, $g_6'(r)$, computed using the reference icosahedra (see text and SI). For the samples at low $\Gamma$, orientational correlations decay rapidly, while at high $\Gamma$ hexagonal order extends across the entire sphere. At intermediate values of $\Gamma$, the absence of well-defined defect clusters (see Fig 2(a)) means that the correlation of the best-fit icosahedron to the orientational order is low, and so many independent configurations must be sampled to required to reveal the equilibrium $g_6'(r)$ curves. Since our experiments only sample a small number of independent configurations, the correlation functions in (b)(i) in the range $85 \lesssim \Gamma \lesssim 110$ sometimes reach negative values at large $r$. The resulting rapid decay in these curves around $\pi/2$ does not reflect equilibrium behavior.
\label{fig:fig3}}
\end{figure*}

In two dimensions, long-range correlations of orientational order, captured by $g_6(\vec{r})\!=\!\langle \psi_6(\vec{x})\psi^*_6(\vec{x}+\vec{r})\rangle_{\vec{x}}$, are the tell-tale sign of crystallinity: however, vector transport on the sphere changes angles and complicates the definition of a global reference coordinate system.
Nevertheless, the icosahedral ordering of defects suggests that it may be possible to detect crystalline order by explicitly referencing this broken symmetry.
We thus define an icosahedral ``net" by rotating an icosahedron so that its vertices are aligned with the positions of the defects (SI \S8).
Projecting the particles onto its faces and unfolding them on a plane reveals the remarkable global orientational coherence of particle configurations with high $\Gamma$ (Fig.~\ref{fig:fig3}a).
We quantify this order using the correlation function
\begin{equation}
\label{g6def}
g'_6(r)\!=\! \langle \psi'_6(\vec{r}_i)\psi'^*_6(\vec{r}_j)\rangle_{|\vec{r}_i-\vec{r}_j|=r}
\end{equation}
where $\psi'_6(r_i)$ is the icosahedrally-referenced value of the bond-orientational order parameter (SI \S8.2).
We note that randomly oriented icosahedral nets, or nets based on polyhedra with different symmetries will not produce such coherence.


The emergence of this new broken symmetry now raises the question of whether the extreme retardation in the growth of $\langle|\psi_6|\rangle$ is a finite-size effect, or reflects the existence of a new order parameter that fundamentally changes the form of the KTHNY transition.
Consequently, to explore the $N \to \infty$ limit, we supplement our simulations of spheres decorated with with $N\!=$1500 particles, with simulations with $N\!=$ 3000, 6000, and 12000 particles~\cite{hoomd1, hoomd2, leimkuhler_efficient_2016}, and evaluate the two order parameters: the icosahedrally referenced 2D bond-orientation of the particles $\Psi_6 = \frac{1}{N}\sum_i{\psi_6'(r_i)}$, and the normalized, three-body, 3D bond-orientation of the defects $\widetilde{W}_6$~\cite{SteinhardtPRB1983},
\begin{equation}
\widetilde{W}_6=-\frac{\sqrt{4199}}{11}\sum^6_{\substack{m_1,m_2,\\m_3=-6}}\left( \begin{array}{ccc} 6 & 6 & 6 \\ m_1 & m_2 & m_3 \end{array} \right) \frac{\rho_{6m_1}\,\rho_{6m_2}\,\rho_{6m_3}}{\big(\sum_m\big|\rho_{6m}\big|^2\big)^{3/2}}
\end{equation}
where the quantity in parenthesis is the Wigner 3-j symbol, $\rho_{6m}\!=\!\sqrt{4\pi}\sum_i Y_{6m}(\theta_i,\phi_i)$, $Y_{6m}$ are the sixth-order spherical harmonics, and ($\phi_i$, $\theta_i$) are the polar coordinates of the $i$-th defect.
This order parameter is particularly sensitive to the presence of icosahedral symmetry, and is normalized so that $\widetilde{W}_6\!=\!1$ for a perfect icosahedron.

\begin{figure*}
\center
\includegraphics[width=8cm]{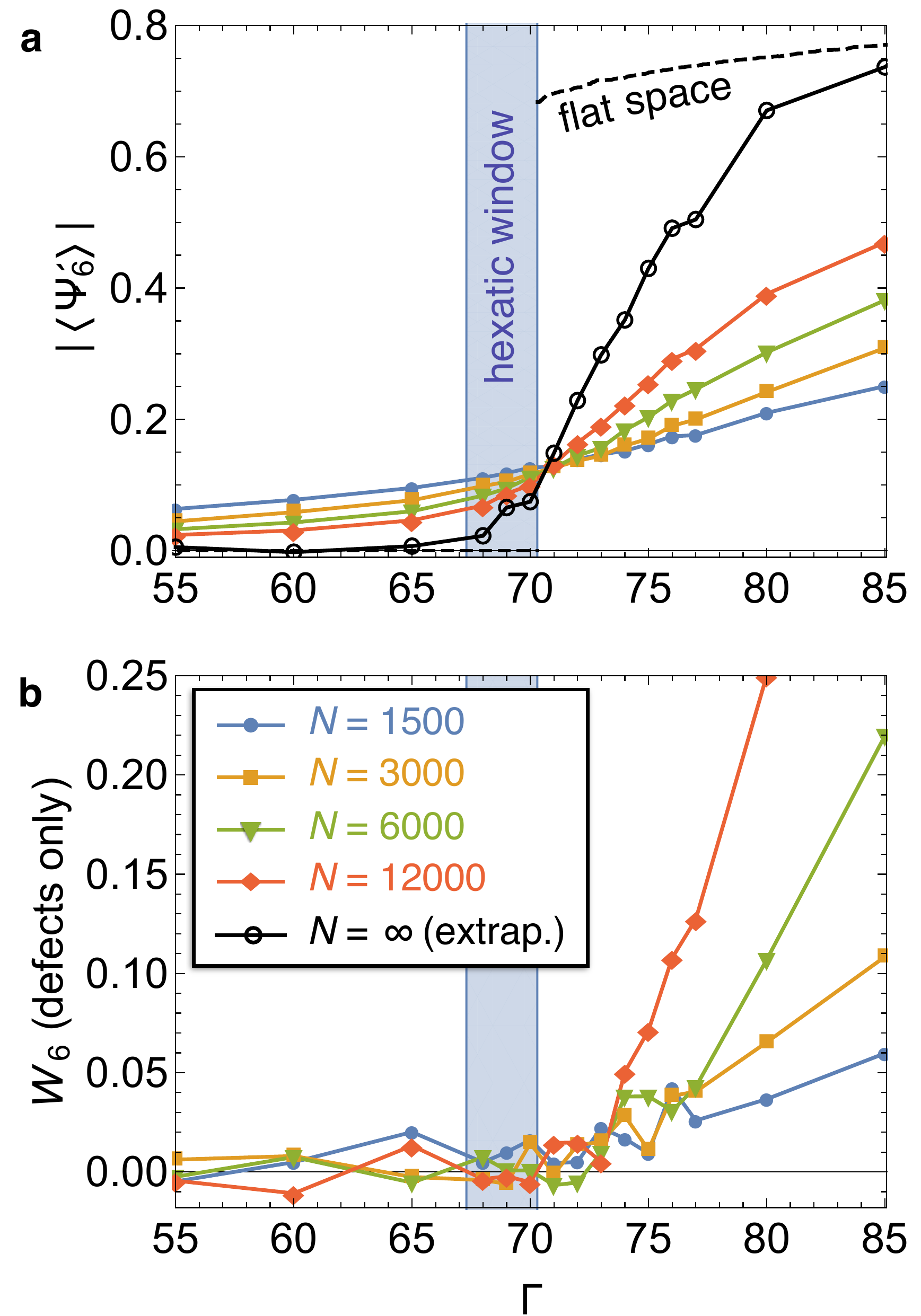}
\caption{\textbf{Icosahedral order and finite size scaling}
(a) Scaling of the normalized, three-body, 3D bond-orientation of the defects, $\widetilde{W}_6$, with $\Gamma$ and system size for simulated particle configurations.
Icosahedral order grows approximately linearly beyond a critical value of $\Gamma$, with a slope that appears to diverge as $N\!\to\!\infty$.
(b) Scaling of the average icosahedrally referenced bond order parameter, $\langle|\Psi_6|^2\rangle^{1/2}$, with $\Gamma$ and system size.
The rate with which this order parameter grows increases with system size, and its $N\!\to\!\infty$ extrapolated values remain positive for all $\Gamma$ that correspond to the flat-space solid.
\label{fig:fig4}}
\end{figure*}

Plots of these quantities show that increasing the system size increases the rate with which orientational and icosahedral order build with increasing $\Gamma$ (Fig.~\ref{fig:fig4}).
A polynomial extrapolation of $\langle|\Psi_6|^2\rangle^{1/2}$ in powers of $1/R \propto 1/\sqrt{N}$ shows that, in the thermodynamic ($N \to\infty$) limit, $\langle|\Psi_6|^2\rangle^{1/2}$ vanishes for $\Gamma<67$,  but remains finite for $\Gamma>70$: coinciding with the 2D liquid and crystal phase boundaries.
Similarly, $\widetilde{W}_6$ is zero for lower $\Gamma$, and begins to increases roughly linearly --- with a slope proportional to between $N^{3/4}$ and $N^1$ --- above a critical threshold.

The elucidation of ordered structure by the unravelling/flattening of an appropriately defect aligned polyhedral net may be useful in other contexts, such as liquid crystalline orientations on curved surfaces (e.g. nematic lines on a baseball, SI \S8.6~\cite{vitelli_nematic_2006}).
Moreover, the segregation of defects to symmetric sites and the concomitant mobility near these sites should prove useful in designing structures where both rigidity and fluidity are desired in specific areas.
This may be of importance in understanding the pervasiveness of icosahedral viral motifs~\cite{CasparKlugQB1962}.

\newpage

\newpage

\newpage

\newpage

\newpage
\bibliographystyle{unsrt}
\bibliography{melting_spheres}

\end{document}